\documentstyle[12pt]{article}
\def\lsim{\raisebox{-4pt}{$\,\stackrel{\textstyle{<}}{\sim}\,$}}
\def\gsim{\raisebox{-4pt}{$\,\stackrel{\textstyle{>}}{\sim}\,$}}
\begin{document}
\begin{flushright}
\baselineskip=12pt
CTP-TAMU-01/97\\
DOE/ER/40717--38\\
ACT-01/97\\
\tt hep-ph/9701264
\end{flushright}

\begin{center}
\vglue 2cm
{\Large\bf Flipped No-scale Supergravity: a Synopsis\\}
\vglue 1cm
{\Large Jorge L. Lopez$^1$ and D.V. Nanopoulos$^{2,3}$\\}
\vglue 0.75cm
\begin{flushleft}
$^1$Bonner Nuclear Lab, Department of Physics, Rice University\\ 6100 Main
Street, Houston, TX 77005, USA\\
$^2$Center for Theoretical Physics, Department of Physics, Texas A\&M
University\\ College Station, TX 77843--4242, USA\\
$^3$Astroparticle Physics Group, Houston Advanced Research Center (HARC)\\
The Mitchell Campus, The Woodlands, TX 77381, USA\\
\end{flushleft}
\end{center}

\vglue 2cm
\begin{abstract}
We discuss the highlights of recent developments in no-scale supergravity
and flipped SU(5), especially in the context of a very light gravitino
scenario that may explain the puzzling CDF $e^+e^-\gamma\gamma+{\rm E_{T,miss}}$ event. We update the status of both subjects and discuss the impact of the latest experimental constraints from LEP and the Tevatron. We also comment on a new form of cold dark matter that may solve this phenomenological `problem' of light gravitino models.
\end{abstract}
\vspace{1cm}
\begin{flushleft}
{\tt lopez@physics.rice.edu}\\
{\tt dimitri@phys.tamu.edu}\\
\medskip
January 1997
\end{flushleft}
\footnotetext{Lecture presented by D.~V.~Nanopoulos at the 34th International
School of Subnuclear Physics ``Effective theories and fundamental interactions", Erice, July 3--12, 1996.}

\newpage

\setcounter{page}{1}
\pagestyle{plain}
\baselineskip=14pt

\section{Introduction}
The next frontier beyond the Standard Model of elementary particle physics is widely expected to contain some sort of low-energy supersymmetry. Indirect evidence in favor of the existence of supersymmetry has been mounting ever since the Tevatron and LEP~1 were commissioned.
This body of evidence includes the convergence of the Standard Model gauge couplings at very high energies in the presence of low-energy supersymmetry, the heaviness of the top quark as required by the radiative electroweak breaking mechanism in supergravity, and the lightness of the Higgs boson as inferred from fits to the high-precision electroweak data. Thus, even though direct manifestations of supersymmetry have not yet been observed (at least not convincingly so), there is plenty of circumstancial evidence that would otherwise be rather difficult to explain, and there certainly are no competing frameworks that even attempt to explain these three facts simultaneously.\footnote{Perhaps the Pope said it best in connection with
his recent endorsement of the theory of evolution: ``The convergence, neither sought nor induced, of results of work done independently one from the other, constitutes in itself a significant argument in favor of this theory."} 

But where are the supersymmetric particles? Why haven't they been found yet,
despite many years of experimental searches? Is absence of evidence evidence
of absence? (as some maintain). Naturalness arguments suggest that the
supersymmetric particles are either `around the corner', or they are not there at all. Moreover, the mass scale of the superpartners is not necessarily related to that of electroweak symmetry breaking ({\em e.g.}, $M_Z,m_t$), but even if
it were (as advanced in `no-scale' models), particle accelerators have
just recently become powerful enough to produce the top quark. Ongoing runs
at LEP~2 and future runs at the Main Injector (starting in 1999) may observe
the first direct evidence of superpartners. However, statistically 
speaking,\footnote{For instance, with a maximum center-of-mass energy of $\sqrt{s}=190\,{\rm GeV}$, LEP~2 hopes to be able to detect the chargino
up to a mass of 95 GeV. Note though that the recent results from LEP~172
have essentially probed 80\% of the total 50 GeV increase in sensitivity over LEP~1. Thus one might say that LEP~2 has left only a 20\% chance of observing
supersymmetry, and with the phase space penalty associated with such relatively
high masses.} it will take the LHC (starting  $\sim$2005) to fully vindicate or falsify the idea of low-energy supersymmetry.

It is important to realize that low-energy supersymmetry {\em per se} will not
be the panacea that will explain the many ad-hoc Standard Model parameters.
In fact, as such it would worsen the situation, as the masses of the many
new particles would become a new set of unknown and unpredicted parameters to
be determined experimentally. Larger symmetries, such as grand unification and
supergravity are expected to correlate the many unknown parameters of both
the Standard Model and its supersymmetric extension, leaving the ultimate
symmetry -- superstrings -- to explain the few leftover fundamental parameters.
However, this program cannot be carried out at all in the absence of low-energy
supersymmetry, in order to deal with the egregious gauge hierarchy problem
which arises because of the disparate mass scales in the theory. Therefore,
one may regard supersymmetry as one of the pillars that would support the eventual `theory of everything', not unlike special relativity and quantum
mechanics supporting the Standard Model. 

\section{The vindication of no-scale supergravity}
Concentrating on the `supersymmetric' parameters, local supersymmetry or supergravity allows one to calculate the masses of the superpartners in terms of two functions: the K\"ahler function $G=K+\ln|W|^2$, where $K$ is the K\"ahler potential and $W$ the superpotential, and the gauge kinetic function $f$. These essentially determine the scalar masses and the gaugino masses respectively. Supergravity has the added bonus of incorporating gravity into
the picture for the first time. A special flavor of supergravity -- no-scale
supergravity \cite{no-scale} -- tackles naturally the problem of the vacuum energy in the presence of supersymmetry breaking. The scalar potential has a zero minimum, which is extended along a flat direction that parametrizes the scale of supersymmetry breaking or the gravitino mass. This picture was proposed back in 1984, and all along has been known to hold well at the tree level, with no clear notion of what would happen at higher orders in perturbation theory. 

Because of its incorporation of gravity, supergravity failed to provide acceptable answers at the Planck scale, where quantum gravity effects are no longer small. This problem is believed to be resolved within the context of
string theory, where the supergravity functions ($G$ and $f$) may be calculated
from first principles, and higher-order corrections may be consistently
obtained in string perturbation theory. In this context Witten \cite{Witten} showed early on that no-scale supergravity emerges in the low-energy limit of weakly-coupled strings. However, string perturbation theory may not be a good approximation to results that are not protected by non-renormalization theorems, such as the K\"ahler potential which so crucially influences the calculation of the vacuum energy. 

Recently there has been a great deal of theoretical progress in the understanding strongly-coupled strings, leading to many connections between previously-thought disconnected pieces of the string puzzle \cite{review}. Moreover, a larger theory -- M-theory -- has been proposed as one that encompasses all of these new properties and describes strongly-coupled strings. For our present purposes, perhaps the most relevant result has been the realization that no-scale supergravity may emerge also in the low-energy limit of M-theory \cite{Horava}, and therefore the no-scale properties may be preserved to all orders in string perturbation theory. In fact, before these results were known, string no-scale supergravity models were studied to lowest order in string perturbation theory \cite{LN}, and found to possess many interesting and intriguing properties \cite{LNZ}. One may now argue that at least the qualitative features of these models ({\em e.g.}, the vanishing of the vacuum energy) might survive to all orders in string perturbation theory. 
 
Having established the central role that no-scale supergravity seems to play,
we should also point out that this framework is more general than a specific K\"ahler function or gauge kinetic function that might realize it.
This generality is reflected in the possible values of the gravitino mass that
are allowable: (i) $m_{3/2}\gg M_W$, (ii) $m_{3/2}\sim M_W$, and (iii) $m_{3/2}\ll M_W$ \cite{no-scale}. The third possibility was first advanced
in Ref.~\cite{EEN} and has motivated the recent surge in interest in light gravitino models within no-scale supergravity.

\section{Whatever happened to Flipped SU(5)?}
Indeed. Let us recount a little history. In 1987 the gauge group SU(5)$\times$U(1) [``flipped SU(5)"] was revived as an interesting and very economical candidate for a unified theory \cite{revitalized}. In the next few years it was realized that this gauge group fit best where others could not fit at all: in string model building \cite{revamped}. The reason being related to the severe lack of large representations in the typical string models of the day (those realizing the gauge group using level-one Kac-Moody algebras), specifically those representations required in the breaking of the unified gauge symmetry down to the Standard Model gauge group. This was a problem for SU(5), SO(10), and $E_6$, but not a problem at all for SU(5)$\times$U(1). Thus, flipped SU(5) appeared singled out.

With the success of the supersymmetric grand unification program, indicating
an apparent convergence of the gauge couplings within the context of the supersymmetric Standard Model at a scale $M_{\rm LEP}\sim10^{16}\,{\rm GeV}$,
it became apparent that something else might be coming to play, as the
string unification scale does not occur (to lowest order) until $M_{\rm string}\sim5\times10^{17}\,{\rm GeV}$. One possibility to obtain a consistent
picture utilizing both mass scales was to try to obtain realistic {\em grand}
unified gauge groups directly from string. This class of models requires
a more sophisticated string construction that is able to realize $k>1$
Kac-Moody algebras. A lot of effort was expended by several groups along these
lines with no clear results \cite{Lykken}. In fact, with the exception of one group \cite{Tye}, efforts to construct a realistic string model based on a grand unified gauge group appeared to have ceased altogether.

Thus, based on present knowledge one is led to conclude that stringy flipped SU(5) is singled out even more singularly now. Furthermore, it has been shown
that flipped SU(5) can make meaningful use of the two mass scales $M_{\rm LEP}$ and $M_{\rm string}$ when the set of string-required hidden-sector fields
come to play in the evolution of the gauge couplings \cite{TwoStep}. 

Flipped SU(5) is also able to solve a relatively new problem in conventional SU(5) grand unification: the prediction for the strong coupling $\alpha_s(M_Z)\gsim0.130$ \cite{Bagger}, which greatly exceeds the experimental world average ($0.118\pm0.003$) \cite{Altarelli}. On the other hand, flipped SU(5) is able to accommodate a range of $\alpha_3(M_Z)$ values, as low as 0.108 \cite{Lowering}. Moreover, values of $\alpha_s(M_Z)$ on the low side imply observable proton decay rates at SuperKamiokande via the $p\to e^+\pi^0$ 
mode \cite{Lowering}.

\section{Flipped no-scale supergravity}
Because of the state of flux that string theory seems to be going through, it is perhaps more advantageous at this point to propose simple unified supergravity models that are heavily inspired from string, rather than 
being rigorously derived (and strongly constrained)  within a framework that
may need to be re-evaluated once M-theory settles down. In this spirit, and
taking the results from the previous two sections into consideration, we would like to concentrate on the generic predictions that might be obtainable in a model with flipped SU(5) gauge symmetry and no-scale supergravity. The latter ingredient we take to imply a very `unified' (and simplified) set of supersymmetry-breaking parameters at the string scale:
\begin{equation}
m_0=A_0=B_0=0\ ,
\label{eq:m0A0B0}
\end{equation}
leaving $m_{1/2}$ as the sole source of supersymmetry breaking. We have
specified neither the Higgs mixing parameter $\mu$, nor the ratio of Higgs-boson vacuum expectation values $\tan\beta$, because these become
derived quantities once the constraints from radiative breaking of the electroweak symmetry are imposed. In this sense we refer to the resulting
superpartner spectrum as a `one-parameter model' \cite{One}, the one free parameter being any superparticle mass scale, such as $m_{1/2}$, or the neutralino mass ($m_\chi$), or the chargino mass ($m_{\chi^\pm}$). (Other
model parameters exist (like the quark masses, etc.), but they are not required to determine the superparticle spectrum.)

\begin{figure}[p]
\vspace{3.5in}
\includegraphics{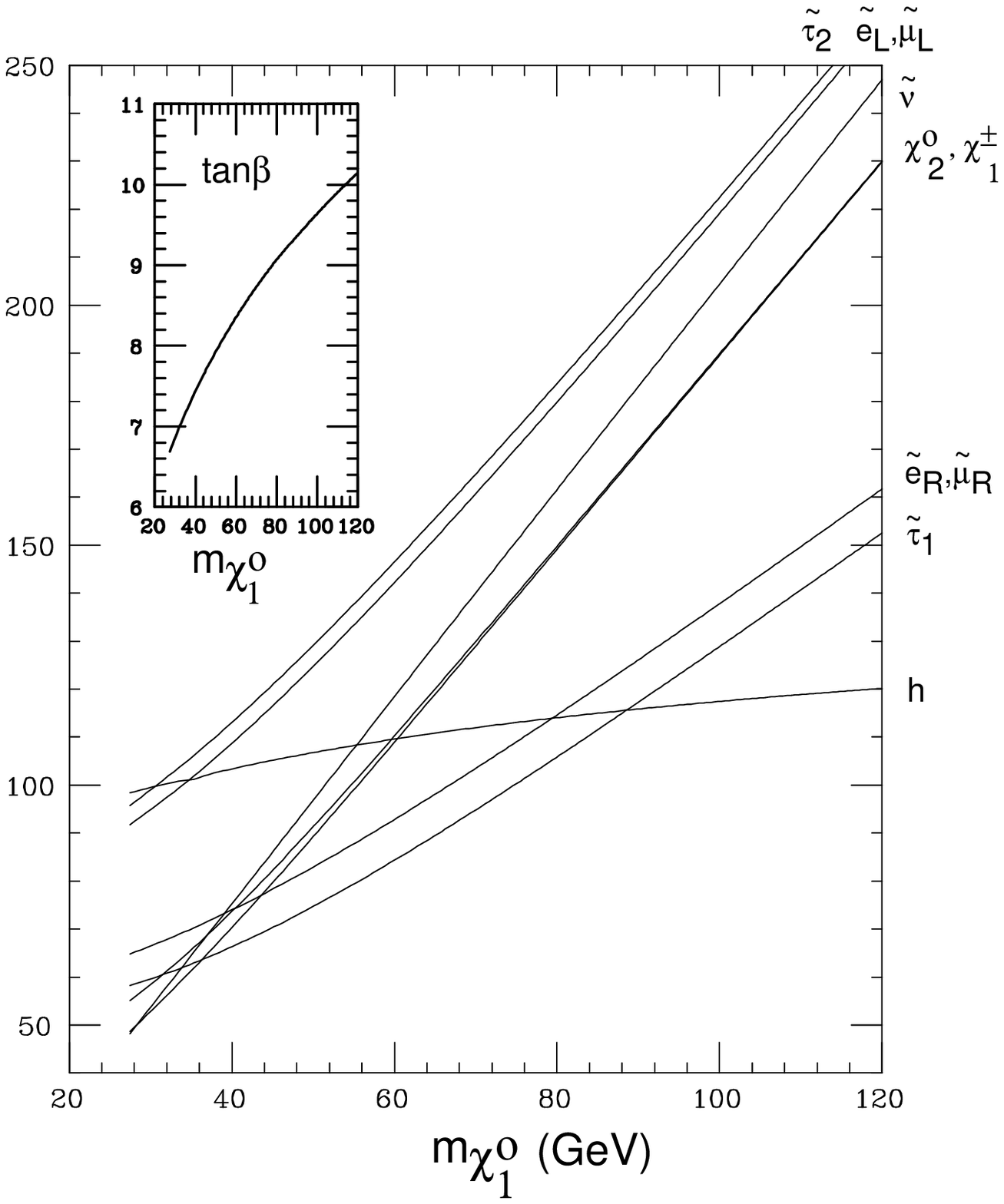}
\vspace{0.7cm}
\caption{The lighter members of the spectrum of our one-parameter model
versus the neutralino mass. All masses in GeV. The inset shows the variation of $\tan\beta$ with $m_{\chi^0_1}$.}
\label{fig:light}
\vspace{3.5in}
\includegraphics{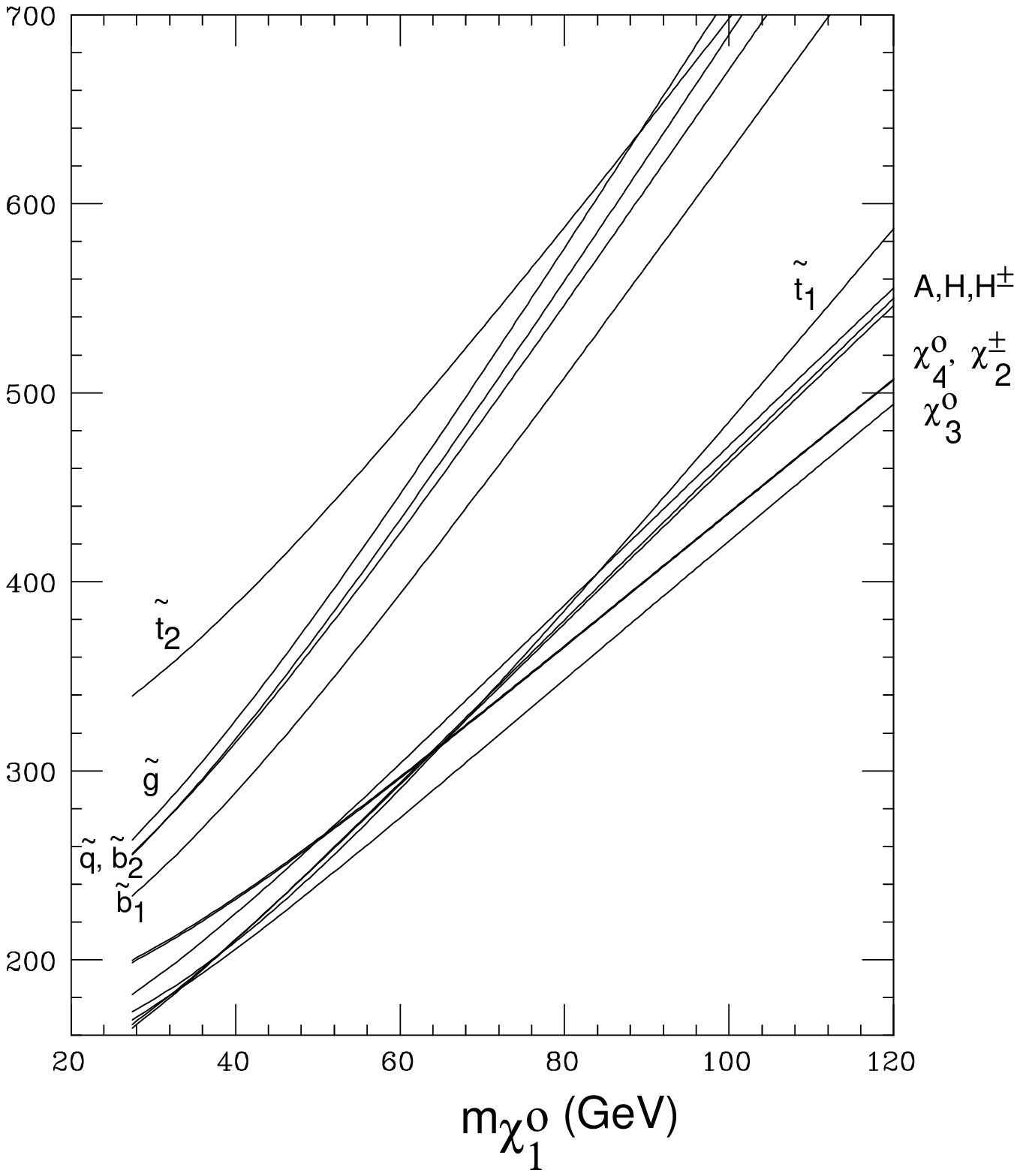}
\vspace{0.7cm}
\caption{The heavier members of the spectrum of our one-parameter model.}
\label{fig:heavy}
\end{figure}

	The calculated spectrum of our one-parameter model \cite{Events} is displayed in Figs.~\ref{fig:light} and \ref{fig:heavy} as a function of the neutralino mass. We note the generally linear increase of all superpartner masses with the neutralino mass. The exception being the lightest Higgs boson mass ($h$), which satisfies $m_h\lsim120\,{\rm GeV}$. In Fig.~\ref{fig:light} we also show the calculated value of $\tan\beta$, which is in the range $7-10$. With such one-parameter spectrum, any observable can be calculated as a function of the neutralino mass. Moreover, experimental limits on such observables can be immediately transformed into limits on $m_\chi$, which then impact the prediction for any other observable. In sum, we have a very predictive and easily falsifiable supersymmetric model.

In order to ascertain the experimental constraints on our one-dimensional parameter space, it is necessary to specify the lightest
supersymmetric particle (LSP), as experimental signatures and sensitivities depend crucially on this choice. In this context there are two choices for the
LSP: the neutralino ($\chi$) or the gravitino ($\widetilde G$). If the gravitino is heavier than the neutralino, then the neutralino is the LSP,
as has been assumed in the majority of phenomenological analyses to date.
This assumption is not unreasonable, as typical supergravity models entail
K\"ahler functions which give $m_0\sim m_{1/2}\sim m_{3/2}$, and the gravitino
is expected to have mass ${\cal O}(M_Z)$. 

In fact, the gravitino may be lighter than the neutralino and the above picture
would effectively still hold, because decays of the neutralino into
the gravitino (which is now the LSP) occur at a very slow rate, {\em i.e.},
outside the detector. This effective neutralino-LSP scenario holds for gravitino masses as low as 250 eV \cite{DineKane}. For lighter gravitinos the neutralino decay rate (which is proportional to $1/m^2_{\tilde G}$) is large enough for the neutralino decay products to become observable in typical detectors, and the gravitino-LSP scenario becomes fully active. This light-gravitino scenario is also not unreasonable, and is predicted in models of gauge-mediated low-energy supersymmetry \cite{DineKane,others} and within no-scale supergravity (as discussed below).

\newpage
\clearpage

The experimental signatures in these two LSP scenarios have some common
aspects, but also some crucial differences. In both cases there is missing
energy in every supersymmetric process, as the LSP escapes detection. However,
in one case the missing energy is carried away by the massive neutralino whereas in the other it is carried away by the essentially massless gravitino.
This difference affects the spectrum of the observable particles, which is softer in the neutralino-LSP case. Thus, missing energy is still the premiere
experimental signature for supersymmetry. A crucial difference between the
two scenarios is the presence of pairs of energetic photons in the gravitino-LSP scenario, and not in the neutralino-LSP scenario.\footnote{Certain neutralino-LSP models \cite{KaneN} do allow photonic signals in restricted regions of parameter space, although
such signals occur only in few specific supersymmetric production processes.}
These photons come from the dominant decay mode $\chi\to\gamma\widetilde G$,
and thus `illuminate' the otherwise `dark' neutralinos. This new signature
increases the efficiency of experimental searches in kinematical configurations
where the observable particles would be too soft to be detectable in the
neutralino-LSP scenario, but which in the gravitino-LSP scenario may always
be tagged by the energetic photons. In practice, one is able to put absolute
lower limits on the sparticle masses which are essentially at the kinematical
limit of the machine (such as LEP). Finally, if the gravitino is sufficiently light, it might be produced directly at {\em e.g.}, electron-positron experiments ({\em i.e.}, $e^+e^-\to\chi\widetilde G$), leading to single-photon signals \cite{Dicus,1gamma}. 

\section{The light gravitino scenario}
The gravitino-LSP scenario has become quite topical recently because of its
ability to naturally explain the intriguing `CDF event' \cite{Park}. As is well known this event, observed by the CDF Collaboration at the Tevatron in $\approx100\,{\rm pb}^{-1}$ of data, is summarized as $p\bar p\to e^+e^-
\gamma\gamma+{\rm E_{T,miss}}$. The momenta of the leptons and photons is
well measured, as is $\rm E_{T,miss}\sim50\,{\rm GeV}$. Two explanations
within the gravitino-LSP scenario have been advanced: selectron pair production
\cite{DineKane,others,Gravitino,Events} and chargino pair production \cite{Gravitino,Events}, with the decay products containing $e^+$, $e^-$,
and neutralinos; the latter decaying into photons plus gravitinos. 

\subsection{How might it be obtained}
In the gravitino-LSP scenario, the main issue is that of {\em decoupling} the breaking of local supersymmetry (parametrized by $m_{3/2}$) from the breaking of global supersymmetry (parametrized by $m_0,m_{1/2}$). This decoupling is achieved {\em naturally} in the context of no-scale supergravity, {\em i.e.},
$m_0=0\cdot m_{3/2}$. Without this (scalar-sector) decoupling, sizeable values of $m_0$ ({\em i.e.}, $m_0\sim M_Z$) cannot be obtained in the light gravitino scenario. The remaining and crucial question is the decoupling in the gaugino sector, which depends on the choice of $f$, at least in traditional supergravity models. The gaugino masses are given by
\begin{equation}
m_{1/2}=m_{3/2}\left({\partial_z f\over 2{\rm Re} f}\right)
\left({\partial_z G \over \partial_{z z^*} G}\right)\ ,
\label{eq:formula}
\end{equation}
where $z$ represents the hidden sector (moduli) fields in the model, and the
gaugino mass universality at the Planck scale is insured by a gauge-group
independent choice for $f$. As remarked above, the usual expressions for $f$ 
({\em e.g.}, in weakly-coupled string models) give $m_{1/2}\sim m_{3/2}$. This result is however avoided by considering the non-minimal choice $f\sim e^{-A z^q}$, where $A,q$ are constants \cite{EEN}. Assuming the standard no-scale expression $G=-3\ln(z+z^*)$, one can then readily show that \cite{EEN}
\begin{equation}
m_{1/2}\sim \left({m_{3/2}\over M}\right)^{1-{2\over3}q} M\ ,
\label{eq:result}
\end{equation}
where $M\approx10^{18}\,{\rm GeV}$ is the rescaled Planck mass.
The phenomenological requirement of $m_{1/2}\sim10^2\,{\rm GeV}$ then implies
${3\over4}\raisebox{-4pt}{$\,\stackrel{\textstyle{>}}{\sim}\,$} q
\raisebox{-4pt}{$\,\stackrel{\textstyle{>}}{\sim}\,$}
{1\over2}$ for $10^{-5}\,{\rm eV}
\raisebox{-4pt}{$\,\stackrel{\textstyle{<}}{\sim}\,$}
m_{3/2}\raisebox{-4pt}{$\,\stackrel{\textstyle{<}}{\sim}\,$}10^3\,{\rm eV}$.
Note that $q={3\over4}$ gives the relation $m_{3/2}\sim
m^2_{1/2}/M\sim10^{-5}\,{\rm eV}$, which was obtained very early on in
Ref.~\cite{BFN} from the perspective of hierarchical supersymmetry breaking in
extended N=8 supergravity. The recent theoretical impetus for supersymmetric
M-theory in 11 dimensions may also lend support to this result, as N=1 in D=11
corresponds to N=8 in D=4.

\subsection{Consistency conditions}
\begin{figure}[p]
\vspace{2.5in}
\includegraphics{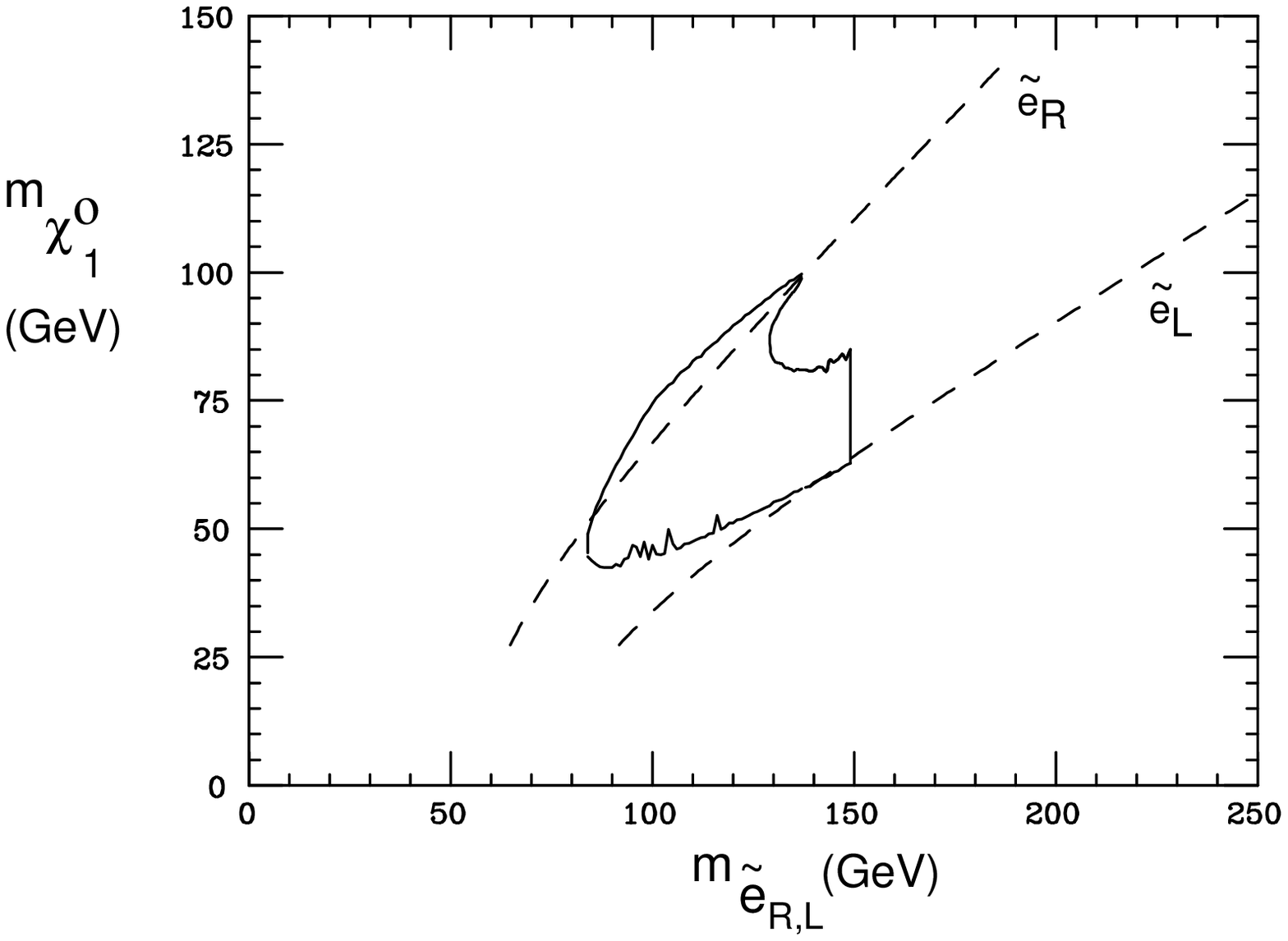}
\vspace{2cm}
\caption{Region consistent with the kinematics of the CDF event when interpreted as selectron pair production. Model predictions are indicated by
dashed lines.}
\label{fig:event1-sel}
\vspace{3in}
\includegraphics{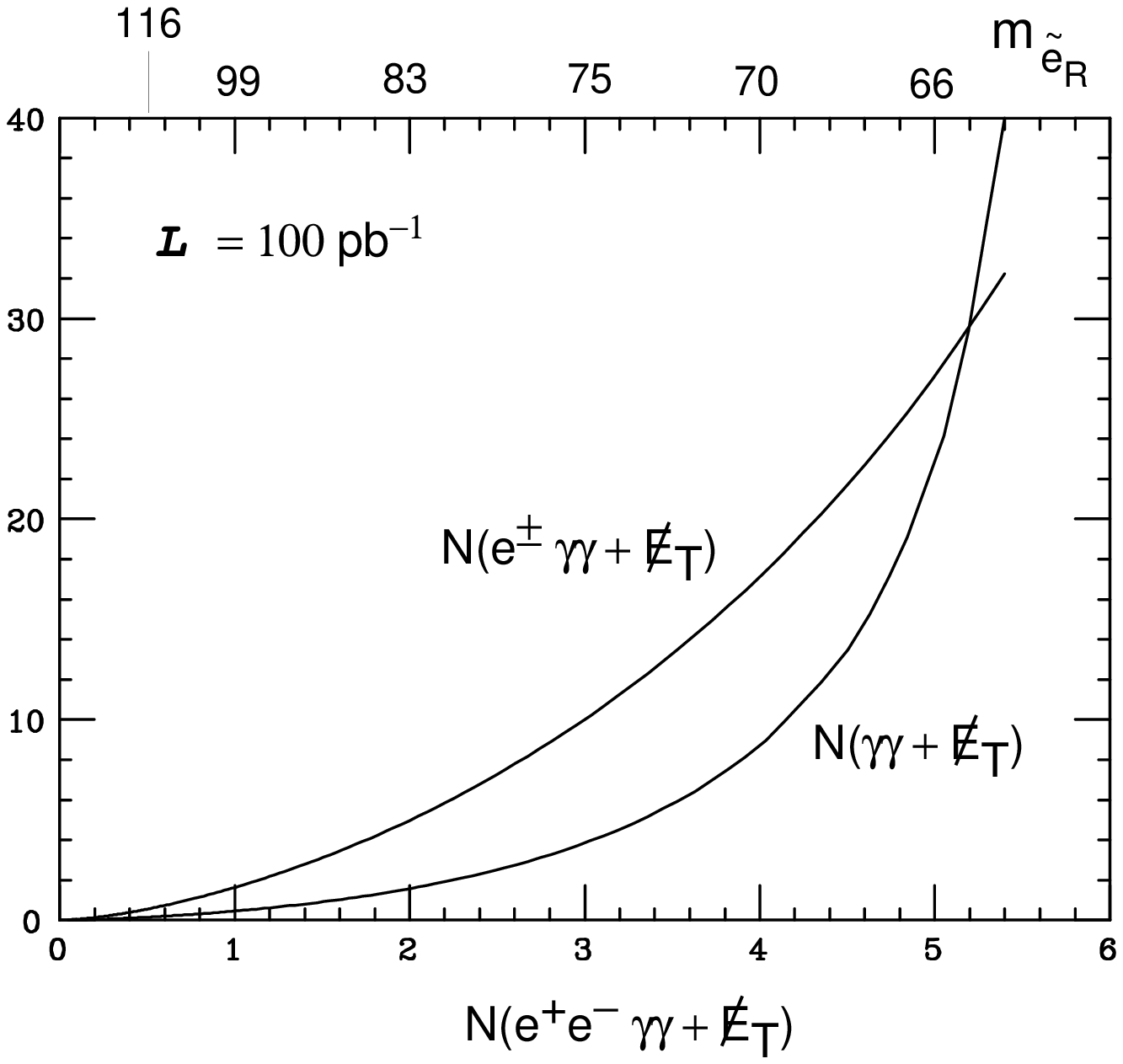}
\vspace{2cm}
\caption{Number of $e^\pm\gamma\gamma+{\rm E_{T,miss}}$
and $\gamma\gamma+{\rm E_{T,miss}}$ events versus the number of
$e^+e^-\gamma\gamma+{\rm E_{T,miss}}$ events expected from slepton production at the Tevatron in ${\cal L}=100\,{\rm pb}^{-1}$ of data. The
top axis shows the corresponding selectron masses ($\tilde e_R$) in GeV.}
\label{fig:1evs2e}
\end{figure}

The two explanations of the CDF event are based on three consistency conditions
that must be satisfied: (i) kinematical consistency, (ii) dynamical consistency, and (iii) overall consistency. As an example let us describe
the simpler case of the selectron interpretation. Kinematical consistency is obtained by determining the region in $(m_{\tilde e},m_\chi)$ space that
yields lepton and photon momenta of the observed values, once the unobserved
gravitino momenta are allowed to vary at random \cite{Gravitino,Events}. The resulting allowed region is depicted in Fig.~\ref{fig:event1-sel}, with the model predictions (from Fig.~\ref{fig:light}) indicated by the dashed lines. The fact that these lines intercept the kinematically-preferred region demonstrates kinematical consistency. 

Dynamical consistency requires a calculation of the selectron pair-production rate and that it be consistent with the one event observed
in $100\,{\rm pb}^{-1}$ of data. Overall consistency is satisfied if any and all related processes occur at their observed levels ({\em i.e.}, zero related
events have been observed). Dynamical and overall consistency are demonstrated
in Fig.~\ref{fig:1evs2e}, where the number of $e^\pm\gamma\gamma+{\rm E_{T,miss}}$ and $\gamma\gamma+{\rm E_{T,miss}}$ events are plotted against
the number of $e^+e^-\gamma\gamma+{\rm E_{T,miss}}$ events, all expected
from slepton production at the Tevatron in ${\cal L}=100\,{\rm pb}^{-1}$ of data. Note that one $e^+e^-\gamma\gamma+{\rm E_{T,miss}}$ event implies much less than one $\gamma\gamma+{\rm E_{T,miss}}$ event and about two 
$e^\pm\gamma\gamma+{\rm E_{T,miss}}$ events. These results are consistent with
present observations. For reference, the top axis in Fig.~\ref{fig:1evs2e} shows the corresponding selectron masses ($\tilde e_R$) in GeV.

\clearpage

\begin{figure}[t]
\vspace{2.5in}
\includegraphics{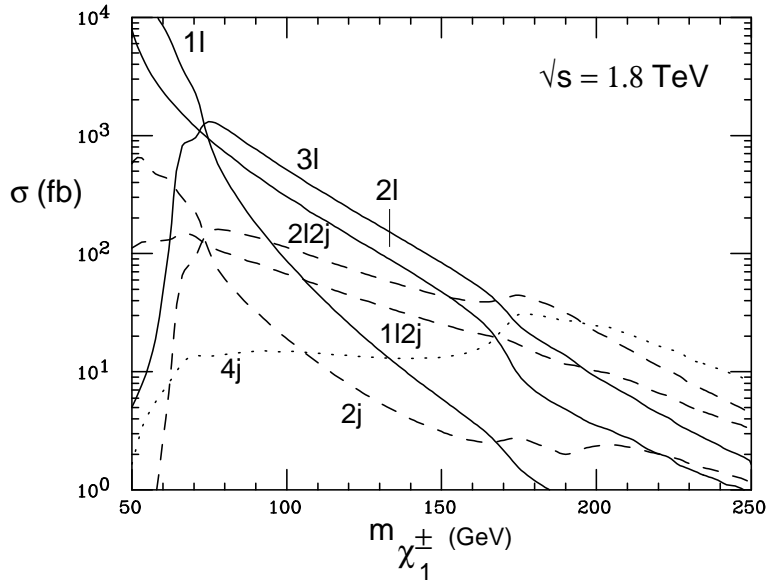}
\vspace{2cm}
\caption{The rates for the various ${\rm n}\ell\,{\rm m}j$ signals (with n charged leptons and m jets) obtained from chargino/neutralino
production at the Tevatron versus the chargino mass. Note that in the
region of interest [$m_{\chi^\pm_1}\approx(100-150)\,{\rm GeV}$] events without jets (solid lines) dominate over events with 2 jets (dashed lines) or events with 4 jets (dotted line).}
\label{fig:signals-cn}
\end{figure}

An analogous but more involved analysis can be carried out in the chargino
interpretation of the CDF event \cite{Events}. The difficulties in this case come from the need to simulate two additional missing particles (the neutrinos in chargino decay) plus the many more chargino decay channels. Moreover, chargino-neutralino production is also important, and it leads to even more
decay channels. Kinematical consistency singles out a region in the
$(m_{\chi^\pm},m_\chi)$ plane, which is intercepted by the predicted line in
the one-parameter model (see Fig.~11 in Ref.~\cite{Events}). Dynamical and overall consistency are demonstrated in Fig.~\ref{fig:signals-cn}. The various curves are obtained by computing the cross sections and branching ratios for the various decay channels. Assuming a 10\% detection efficiency, one `$2\ell$' event in $100\,{\rm pb}^{-1}$ of data appears consistent with the range $m_{\chi^\pm}\approx(100-150)\,{\rm GeV}$. Most interestingly, in this range one sees that the events with no jets dominate over those with jets. This is a feature of the data that is naturally explained in this model by the specific branching ratios, but that fails to be explained in alternative light-gravitino models \cite{BCPT}.

\begin{figure}[t]
\vspace{2.5in}
\includegraphics{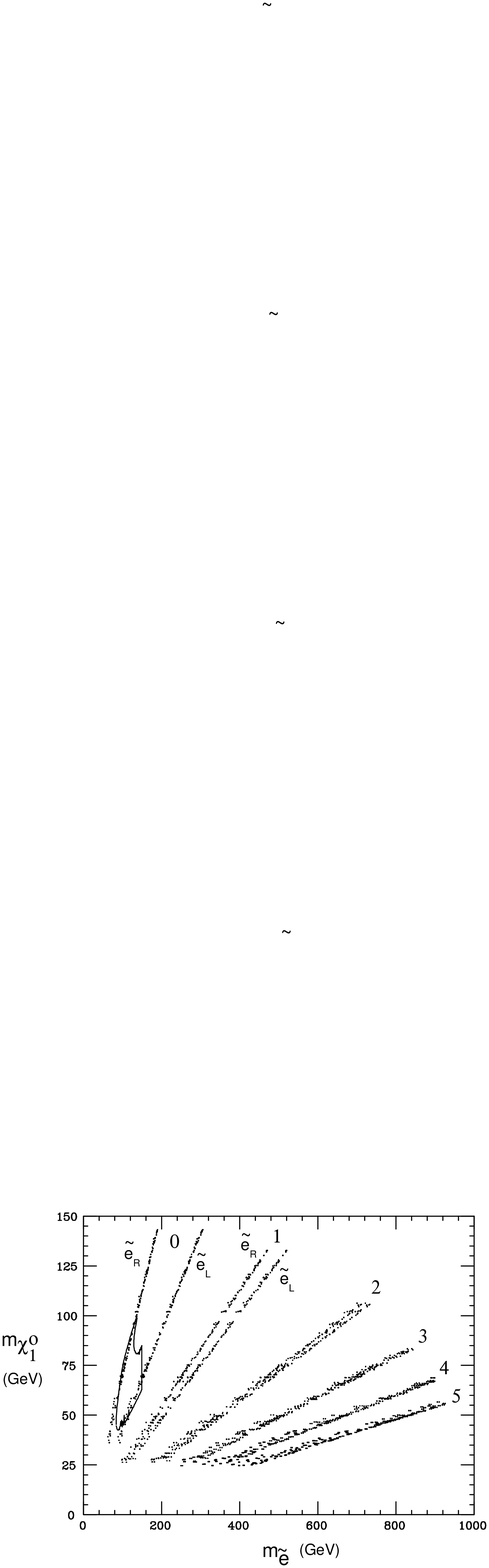}
\vspace{2cm}
\caption{Calculated distribution of selectron ($\tilde e$) and lightest
neutralino ($\chi^0_1$) masses in generic supergravity models for
fixed values of the ratio $\xi_0=m_0/m_{1/2}=0,1,2,3,4,5$; and varying values
of $\{m_{1/2}, \tan\beta,A_0\}$.}
\label{fig:lspsel}
\end{figure}

\subsection{The role of flipped SU(5)}
Let us now reflect on the kinematically preferred region (at least in the
selectron interpretation) from the supergravity perspective. A scan of the four-dimensional parameter space $\{m_{1/2},m_0,A_0,\tan\beta\}$ allows one to calculate $m_{\tilde e}$ and $m_\chi$ for, say, fixed values of $\xi_0=m_0/m_{1/2}$ and floating values of the other three parameters. The resulting scatter plot, shown in Fig.~\ref{fig:lspsel}, reveals that $\xi_0\approx0$ is required to match the kinematical data from the CDF event \cite{Events}. This result is expected because the preferred selectron masses are quite light. In the context of SU(5) supergravity grand unification this requirement on $\xi_0$ leads to proton decay rates via dimension-five operators ($p\to\bar\nu K^+$) that are inconsistent with experimental limits (which demand $\xi_0\gg1$ \cite{pdecay}). So, proton decay and the light gravitino explanation of the CDF event are incompatible within minimal SU(5) GUTs. Flipped SU(5), on the other hand, is perfectly compatible with such low values of $\xi_0$ because dimension-five proton decay operators are naturally suppressed by the `flipped' gauge structure \cite{revitalized}. Of course, $\xi_0=0$ is derivable from the SU(N,1)/U(1) no-scale supergravity 
structure \cite{no-scale}.

\begin{figure}[t]
\vspace{3in}
\includegraphics{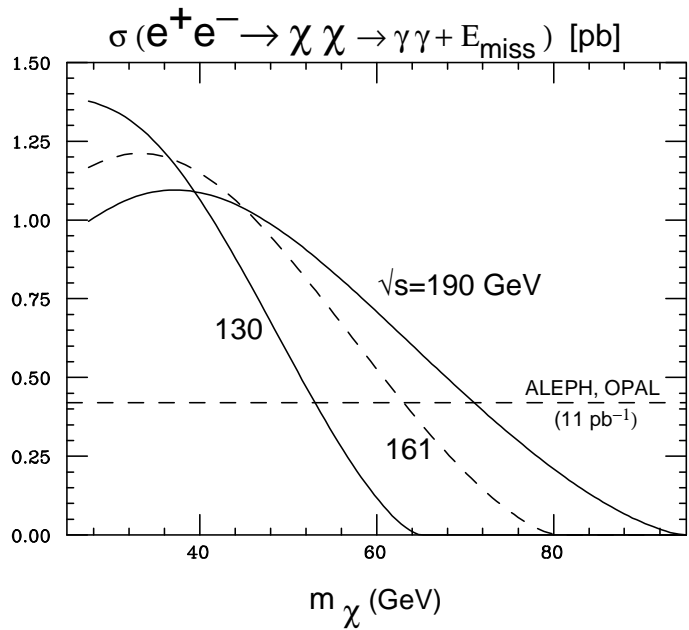}
\vspace{2cm}
\caption{Diphoton cross section versus neutralino mass at various LEP center-of-mass energies in no-scale supergravity with a light gravitino. The preliminary ALEPH and OPAL upper limits obtained at LEP161 are indicated.}
\label{fig:N1N1-updated}
\end{figure}

\section{Expectations at LEP and the Tevatron}
\subsection{LEP}
The fact that the neutralino is unstable and decays within the detector into
a photon and a gravitino makes the channel 
\begin{equation}
e^+e^-\to\chi\chi\to\gamma\gamma+{\rm E_{miss}}
\label{eq:2gamma}
\end{equation}
``visible". This channel also has the largest reach into parameter space,\footnote{Excluding the $e^+e^-\to\chi\widetilde G$ channel, which has
an ever larger reach, but whose rate is very sensitive to the gravitino mass.}
as $\chi$ is the next-to-lightest supersymmetric particle. The cross section
as a function of the neutralino mass is shown in Fig.~\ref{fig:N1N1-updated}.
Note that present data require $m_\chi\gsim65\,{\rm GeV}$. 

There has been speculation that some LEP diphoton events might in fact be
attributable to the process in Eq.~(\ref{eq:2gamma}) \cite{diphotons}. Such `candidate' events have turned up at the rate of one or two per run and have been reported by a few LEP Collaborations. However, a remaining shadow of doubt remains because of their possible explanation as background events from 
$e^+e^-\to\gamma\gamma Z\to\gamma\gamma\nu\bar\nu$, which have a missing invariant mass distribution that peaks sharply near $M_{\rm miss}\approx M_Z$,
and which are also expected to occur at a rate of one or two per run per
experiment. The selected candidate events have $M_{\rm miss}\approx100,{\rm GeV}$, and thus it is not clear what their origin might be.

The LEP diphoton events can be scrutinized more deeply by assuming that they
indeed come from neutralino pair production, and then conducting a kinematical
analysis analogous to that used in the case of the CDF event \cite{Events,diphotons}. In this case the number of kinematical constraints is larger (well known total momentum and energy) and allows one to obtain a range of allowed neutralino masses consistent with the diphoton kinematics. These analyses indicate that most of the candidate events require neutralino masses otherwise excluded by direct searches ({\em i.e.} $m_\chi<65\,{\rm GeV}$). However, one must keep in mind that particles produced near the kinematical limit may be off shell, invalidating the analysis, and effectively widening the allowed neutralino mass interval. More data at higher center-of-mass energies are required to reach a definitive conclusion.

\subsection{Tevatron}
Searches for `CDF-like' events at the Tevatron have been conducted, although
not necessarily on a channel-by-channel basis but rather on an inclusive
basis, where all events with two photons plus missing energy are analyzed.
A preliminary analysis from CDF has been released \cite{Toback}, and more recently a full analysis from D0 \cite{D0gamma}. The D0 analysis has been tailored to an alternative explanation of the event in the neutralino-LSP scenario. Nonetheless, limits have been given for more general underlying scenarios: 
\begin{equation}
\sigma\cdot B(p\bar p\to\gamma\gamma+{\rm E_{T,miss}}+X)<185\,{\rm fb}
\label{eq:D0limit}
\end{equation}
with the following restrictions
\begin{equation}
E^\gamma_T>12\,{\rm GeV},\quad |\eta^\gamma|<1.1,\quad {\rm and}\quad
{\rm E_{T,miss}}>25\,{\rm GeV}\ .
\label{eq:restrictions}
\end{equation}
It is not easy to use this limit to constrain the model at hand. One would
need to simulate all of the possible decay channels individually and then
collect the number of events that passed all cuts in Eq.~(\ref{eq:restrictions}). Instead, one can obtain approximate constraints
on the parameter space by assuming \cite{D0gamma} that 25\%--50\% of the events would pass these cuts, thus imposing an upper limit on the total (no cuts) inclusive diphoton cross section which is a factor of 2 to 4 larger than that in Eq.~(\ref{eq:D0limit}): (370--740) fb. Adding up all of the curves in
Fig.~\ref{fig:signals-cn}, one finds that 
\begin{equation}
m_{\chi^\pm}\gsim(110-130)\,{\rm GeV},\qquad m_\chi\gsim(60-70)\,{\rm GeV}
\label{eq:newlimits}
\end{equation}
appear required. These limits are still perfectly consistent with the selectron and chargino interpretations of the event. Moreover, the limit on the neutralino mass is remarkably close to that presently obtainable at LEP via
neutralino pair-production searches.

\section{A new source of dark matter}
Besides the ubiquitous photonic signature of gravitino-LSP models, perhaps
the other notable difference with traditional neutralino-LSP models is the
lack of an obvious candidate for the dark matter in the Universe. The relic
gravitinos themselves, as is well known, for $m_{3/2}\sim1\,{\rm KeV}$  constitute a form of `warm' dark matter with a behavior similar to that of cold dark matter. The non-thermal gravitinos from $\chi$ decay do not disturb big bang nucleosynthesis, and may constitute a form of hot dark matter \cite{BMY}, although with small abundance. Other forms of dark matter, such as metastable hidden sector matter fields (cryptons) \cite{cryptons} and a cosmological constant \cite{Age}, may need to be considered as well.

There is however, another candidate for cold dark matter in light gravitino
scenarios that has been overlooked in the recent revival of this idea, namely
the coherent oscillations of axion-like particles in the model \cite{EENII}.
These fields are essentially the real and imaginary components of the hidden
sector field $z$ that appears in Eq.~(\ref{eq:formula}), more commonly
referred to in the literature by their scalar ($S$) and pseudo-scalar ($P$) linear combinations. If these axion-like fields are as light as the gravitino, they contribute to supernova and stellar energy loss \cite{EENII,GMR} and
constrain the possible values of the gravitino mass.

The values of $m_{3/2}$ and $m_{S/P}$ depend only on $m_{1/2}$ and the choice of $p=1/(1-{2\over3}q)$ in Eq.~(\ref{eq:result}): $m_{3/2}\sim m^p_{1/2}$ and $m_{S/P}\sim\Lambda^2_{\rm QCD}/m^{p-1}_{1/2}$ \cite{EENII}. Preliminary studies show that there is a range of $p$ values for which (i) all astrophysical constraints on the gravitino and $S/P$ particles are satisfied, (ii) the axion-like $S/P$ particles contribute a significant component to the cold dark matter in the Universe, (iii) the neutralinos decay inside the detectors in collider experiments, and (iv) all laboratory constraints on the light gravitinos are satified \cite{S/P}. Note that this is not
the case in alternative gravitino-LSP models \cite{DineKane,others}, which
have been recently shown to require much heavier gravitino
massed ($m_{3/2}\sim100{\rm KeV}$) leading to much larger than closure
gravitino dark matter abundances which need to be diluted somehow \cite{GMM}. 

\section{Conclusions}
We have provided a synopsis of recent developments in no-scale supergravity
and the everpresent central role that it appears to play in string theory
and its modern generalizations. We have also updated the status of flipped
SU(5) and the demise of its traditional GUT alternatives in string model
building. A model based on no-scale supergravity and flipped SU(5) may
explain the puzzling CDF event and have many more falsifiable predictions.
Ongoing runs at LEP 2 and future runs at the Main Injector should support
this picture. Finally, we have discussed a solution to a poignant `problem'
of gravitino-LSP models -- their lack of a clear cold dark matter candidate.

\section*{Acknowledgments}
The work of J.~L. has been supported in part by DOE grant DE-FG05-93-ER-40717 and that of D.V.N. by DOE grant DE-FG05-91-ER-40633.

\end{document}